\documentclass[twocolumn,showpacs,preprintnumbers,amsmath,amssymb]{revtex4}

\usepackage{graphicx}
\usepackage{dcolumn}
\usepackage{bm}
\usepackage{textcomp}
\usepackage{wasysym}
\usepackage{stmaryrd}

\def \ee{\end{equation}}
\def \be{\begin{equation}}

\preprint{}

\begin{document}

\title{(2+3) dimensional geometrical dual of the complex Klein-Gordon equation}

\author{Benjamin Koch}
 \affiliation{
 Institut f\"{u}r Theoretische Physik, Johann Wolfgang Goethe -
Universit\"{a}t,\\
D--60438 Frankfurt am Main, Germany\\
}

\date{\today}

\begin{abstract}
In this paper it is shown that an equivalent to
the complex Klein-Gordon equation can be obtained from
the (2+3) dimensional Einstein equations coupled to
a conserved energy momentum tensor.
In an explicit toy model
we give matching conditions for what corresponds to
the phase, the amplitude, and the mass of the complex wave function.
\end{abstract}

\maketitle

The purpose of this paper is to show that
one can find an analog to the complex Klein-Gordon
equation in curved space-time by considering matter which is moving
in an higher dimensional space-time.
This is motivated on the one hand 
by the fact that there are alternative approaches 
to quantum mechanics.
Early such attempts have been discussed within the so called
Bohmian mechanics \cite{Bohm:1951,Bohm:1951b}.
Similarities between a higher dimensional wave equation
and the non-relativistic quantum theory have been pointed
out in \cite{Klein:1926a}.
It was also suggested that
quantum field theory might emerge from a chaotic classical theory
with friction \cite{Hooft:1999gk,Biro:2001dh}.
For an overview see \cite{Nikolic:2006az}.
On the other hand a further motivation comes from
the existing ideas on an
additional time dimension. The consequences of
more than one time dimensions were 
studied in
\cite{Dvali:1999hn,Bars:2000qm,Bars:2001ma,Villanueva:2005up,Bars:2006dy}.
Further, an additional time dimension was used to
interpret geodesics in four dimensions as null paths
in the higher dimensions \cite{Seahra:2001bx}.
There is also an intersection of both motivations.
Some papers relate a quantum field
theorie on the horizon of a black hole to a corresponding
classical theory in the higher dimensional spacetime
\cite{Hooft:1993gx,Susskind:1994vu}.
Similar effects are found in the context of supergravity and string theory 
involving a holographic principle and extra time dimensions 
\cite{Maldacena:1997re,Witten:1998qj,Hull:1999mr}.
In a purely conceptual discussion
an extra time dimension is suggested as an
alternative to quantization \cite{Weinstein:2007}.
In a stochastic approach \cite{Damgaard:1987rr}, the classical
Langevin equation for an auxiliary time coordinate $\tilde{t}$
is used in order to obtain Euclidian quantum field theory
in the limit $\tilde{t}\rightarrow \infty$.
All those ideas can be seen as motivation for
 this paper, where the structure of the complex Klein-Gordon
equation is found from
classical differential geometry with one additional time dimension.

The motivation of our approach is the idea  that the
movement of a classical particle
with respect to an additional but unobservable
time dimension $\bar{t}$ could produce effects 
which are usually described by quantum mechanics.
As first application of this idea we will
derive the structure of the relativistic complex Klein-Gordon equation 
from a classical theory with one additional dimension.
%
\section{The complex Klein-Gordon equation
in curved spacetime}

The
relativistic Klein-Gordon equation in curved spacetime \cite{Birrell} is
\be\label{eq_KG0}
\frac{1}{\sqrt{-\hat{g}}}\partial_\mu
\left(\sqrt{-\hat{g}} \hat{g}^{\mu \nu}\partial_\nu \Phi(x,t)\right)=-\frac{m^2}{\hbar^2} \Phi(x,t)\quad,
\ee
where $\hat{g}$ stands for the determinant of the metric $\hat{g}_{\mu\nu}$.
This equation for the complex wave function $\Phi$ can be expressed
in terms of the real
function $S_Q(x,t)$ and the real function $\rho(x,t)$ by defining
$\Phi(x,t)=\sqrt{\rho} \exp(i S_Q/\hbar)$.
This gives two coupled differential equations
\begin{eqnarray}\label{eq_KG1}
\boxempty\sqrt{\rho}
&=& \frac{\sqrt{\rho}}{\hbar^2} 
\left( (\partial^\mu S_Q)(\partial_\mu S_Q) -m^2\right)\quad, \\ 
\label{eq_KG2}
0&=&
\partial_\mu \left(\sqrt{-\hat{g}}\rho \hat{g}^{\mu \nu}(\partial^\nu S_Q) \right)\quad.
\end{eqnarray}
In the first equation we used the abbreviation $\boxempty \sqrt{\rho} \equiv \partial_\mu
\left(\sqrt{-\hat{g}}\hat{g}^{\mu \nu} \partial_\nu \sqrt{\rho} \right)/\sqrt{-\hat{g}}$
for the covariant d'Alembert operator.

The redefinition of the complex wave function 
in terms of the real functions $S_Q$ and $\rho$ does not change the meaning or the interpretation of Eq.~(\ref{eq_KG0}).
The function $S_Q$ is usually understood as the phase of the
quantum wave function. 
%
\section{General relativity with one additional time dimension}

In this section we will consider the geometric structure of 
Einsteins equations with one additional coordinate $\bar{t}$.
A simple ansatz for the
metric in $2+3$ dimensions will be made. 
We  then show that
it is possible to choose the energy-momentum tensor $T_{AB}$
in such a way that the classical equations of motion
in the higher dimensional theory (after the
additional coordinate $\bar{t}$ is integrated out) 
correspond to the relativistic quantum
equation for a spinless particle.
We use the coordinate notation $x_A=(\bar{t},x_\mu)=(\bar{t},t,\vec{x})$,
where capital Latin indices $A$ 
run from $0$ to $4$ and 
Greek indices run from $1$ to $4$.
As starting point we take the
Einstein equations in (2+3) dimensions
\begin{eqnarray}\label{eq_Einstein}
R_{AB}-\frac{1}{2} g_{AB} R
=G_D T_{AB}
-\frac{1}{2} g_{AB}
\Lambda \quad, 
\end{eqnarray}
where $R_{AB}$ is the higher dimensional Ricci tensor, $R$
is it's contraction, and $G_D$ is the higher dimensional
gravitational coupling constant.
The term $\Lambda$ is directly connected 
to the cosmological constant $\Lambda_4$.
Our ansatz for the higher dimensional metric is
\be\label{Ansatz0}
g_{AB}dx^A dx^B=
\bar{\alpha}^2(\bar{t}) \rho(t,\vec{x})d\bar{t}^2+
g_{\mu \nu}(\bar{t},x_\alpha) dx^\mu dx^\nu.
\ee
Later, the function $\bar{\alpha}$ will be separated 
into a constant plus a small $\bar{t}$ dependent part
\be
\bar{\alpha}(\bar{t})=\alpha_0+\epsilon_0 \bar{\omega}(\bar{t})\quad .
\ee
Note that this metric is similar to the Kaluza-Klein
metric for a vanishing electromagnetic field $A_\mu=0$ 
\cite{Kaluza:1921,Klein:1926a,Klein:1926b,Wesson:2000}.
However, in contrast to the Kaluza-Klein approach the
energy momentum Tensor $T_{AB}$ and
$\Lambda$ are not assumed to be identically zero and
all the functions have real values only.
An other difference is that
the functions in the metric ($\bar{\alpha}^2 \rho$ and $g_{\mu \nu}$)
and the energy momentum tensor $T_{AB}$ are allowed
to have a explicit $\bar{t}$ dependence.
With the metric~(\ref{Ansatz0}) the twenty-five coupled
differential equations~(\ref{eq_Einstein}) read
\begin{eqnarray}\nonumber
 \left(
-\bar{\alpha}^2 \sqrt{\rho} \boxempty \sqrt{\rho}
-\frac{(g^{\lambda \beta}\dot{g}_{\lambda \beta})_{,0}}{2}
+\frac{\dot{\bar{\alpha}}g^{\lambda \beta}\dot{g}_{\lambda \beta}}{2 \bar{\alpha}}
\quad\quad\right.\\ \nonumber
\left.
-\frac{g^{\mu \beta} g^{\lambda \sigma} \dot{g}_{\lambda \beta}\dot{g}_{\mu \sigma}}{4}\right)\\\label{loong1}
= \left[G_D \left(T_{00}-\frac{\bar{\alpha}^2
\rho}{3}T^A_A\right)
+\frac{\bar{\alpha}^2 \rho}{3}\Lambda\right] \quad,
\end{eqnarray}
\begin{eqnarray}\nonumber
 \left(
\frac{g^{\lambda \beta}}{4\rho}(\dot{g}_{\lambda \beta}\rho_{,\delta\,}
- \rho_{,\beta\,}\dot{g}_{\delta \lambda})
+ \frac{\partial_\lambda(g^{\lambda \mu}\dot{g}_{\mu \delta})}{2}\right. 
\quad \quad \quad \\ \nonumber
\left.- \frac{\partial_\delta(g^{\lambda \mu}\dot{g}_{\lambda \mu })}{2}
+\frac{g^{\lambda \sigma} g^{\mu \beta} \dot{g}_{\sigma \delta} g_{\mu \beta,\lambda}}{4}
+\frac{\dot{g}^{\mu \beta} g_{\mu \beta,\delta}}{4}\right)\\ \label{loong2}
=G_D T_{0 \delta}\quad,\,\;
\end{eqnarray}
\begin{eqnarray}\nonumber
\left[\hat{R}_{\delta \beta} -
\frac{(\sqrt{\rho})_{;\delta;\beta}}{\sqrt{\rho}}
+\frac{1}{2 \bar{\alpha}^2 \rho}\left(\frac{\dot{\bar{\alpha}}\dot{g}_{\delta \beta}}{\bar{\alpha}}
-\ddot{g}_{\delta \beta}
\right.\right.\\ \nonumber
\left.\left.
+g^{\lambda \mu}\dot{g}_{\delta \lambda}\dot{g}_{\beta \mu}
-\frac{g^{\mu \nu}\dot{g}_{\mu \nu}\dot{g}_{\delta \beta}}{2}\right)\right]\\ \label{loong3}
=\left[G_D \left( T_{\delta \beta}-\frac{g_{\delta \beta}}{3}
T^A_A\right)+
\frac{g_{\delta \beta}}{3}\Lambda\right]\quad.
\end{eqnarray}
Here, we denoted a derivative with respect to $\bar{t}$ as
$\partial_0 X\equiv \dot{X}$, the covariant derivative in the
four dimensional subspace as $\nabla_\mu X\equiv X_{ ;\mu}$, and
$\boxempty \sqrt{\rho}=(\sqrt{\rho})^{;\mu}_{\,;\mu}$.
$\hat{R}_{\delta \beta}$ is the normal four dimensional form of the Ricci tensor
which only contains derivatives of $g_{\mu \nu}$ with respect to $x_{\mu}$. 
The Hamilton-Jakobi definition of the
energy momentum tensor of a free particle
in curved spacetime is
\begin{eqnarray}\label{eq_TAB}
T^A_{\;\;B}&=&(\partial^A S_H) (\partial_B S_H),
\quad 
\mbox{with}\quad \\ \nonumber 
T^A_{\;\;A} &=&\left((\partial^0 S_H) (\partial_0 S_H)+
(\partial^\mu S_H) (\partial_\mu S_H)\right)
\quad,
\end{eqnarray}
where $S_H$ is the density of Hamilton's principal
function \cite{Hamilton:1833,Landau:1975}.
A priori, the $\bar{t}$ dependence of $S_H$ is not known
Therefore, we make a separation ansatz where 
$S_H$ can be decomposed into a classical part 
$\tilde{S}$ which
only depends on the observable coordinates $t,\vec{x}$
and a part $B(\bar{t}) \sqrt{\rho}$ 
which depends on all coordinates $\bar{t},t,\vec{x}$:
\be\label{eq_SH}
S_H(\bar{t},t,\vec{x})=\epsilon_1 \sqrt{\rho}B(\bar{t})+\tilde{S}(t,\vec{x})\quad,
\ee 
with $\partial_0 B(\bar{t})\equiv \beta(\bar{t})$.
Instead of making the heroic but vain attempt to 
solve all those coupled equations simultaneously we restrain 
ourselves to a simpler scenario.
In this scenario we take the trace  of the
higher dimensional Einstein equations and integrate
over additional dimension $\bar{t}$
\be\label{eq_EinsteinScal}
\int d\bar{t}\,R=\frac{-1}{3}\int d\bar{t}\,(2 G_D T^A_{\;A}-5\Lambda)\quad.
\ee
Using the definitions~(\ref{Ansatz0}, \ref{eq_TAB}),
and after a multiplication with $-\sqrt{\rho}/2$
the scalar equation~(\ref{eq_EinsteinScal})
reads
\begin{eqnarray}\label{eq_dgl2c}
 \boxempty \sqrt{\rho} \int d\bar{t} = 
 \sqrt{\rho}
 \left[\frac{G_D}{3}\int d\bar{t} \,(\partial^\mu S_H)(\partial_\mu S_H)
 \right. \quad \quad \quad  \\ \nonumber\left.
- \left( \frac{5}{6}\Lambda  \int d\bar{t} \,
- \frac{ \int d\bar{t} \, \hat{R}^\alpha_\alpha}{2}  
 -\frac{G_D \epsilon_1^2}{3} 
 \int d\bar{t} \,\frac{\beta^2}{\bar{\alpha}^2}
 \right)\right] \\ \nonumber -
\int d\bar{t}\,\frac{1}{\bar{\alpha}}\partial_0 \frac{g^{\alpha \beta}\dot{g}_{\alpha \beta}}{\sqrt{\rho} \bar{\alpha}}
+\frac{\int d\bar{t}}{2\sqrt{\rho}}   
\left( \frac{P^{\alpha \beta}P_{\alpha \beta}}{ \bar{\alpha}^2} -\frac{(P^\alpha_\alpha)^2}{9 \bar{\alpha}^2} \right).
\end{eqnarray}
The source term on the right hand side
contains the tensor 
$P_{\mu \nu}=(\dot{g}_{\mu \nu}-g_{\mu \nu} g^{\alpha \beta}\dot{g}_{\alpha \beta})/2$.
Apart from the last two terms this equation has
always the same $\rho$ dependence as Eq.~(\ref{eq_KG1}), which justifies the
ansatz~(\ref{eq_SH}).
Those last terms have the form of a source, which
indicates that a $\bar{t}$ dependence of $g_{\mu \nu}$
can lead to particle production.
This result is not too surprising because it is known
that already a normal time dependence of the metric
can lead to particle production \cite{Bernard:1977pq}.
However, we are primarily interested in problems
with a seemingly constant (with respect to $\bar{t}$) four dimensional spacetime.
Therefore, we decompose the four dimensional submetric $g_{\mu \nu}$
into a $\bar{t}$ independent part $\hat{g}_{\mu \nu}$ 
and a traceless $\bar{t}$-variable part 
$\gamma_{\mu \nu}(\bar{t},x^\alpha)$
\be\label{eq_gExpand}
g_{\mu \nu}(\bar{t},x^\alpha)=\hat{g}_{\mu \nu} + \epsilon_2^2 \gamma_{\mu \nu}(\bar{t},x^\alpha)
\quad.
\ee
Now we assume that perturbations due to $\bar{t}$
are small and expand Eq.~(\ref{eq_dgl2c}) 
to lowest order in $\left. \epsilon_i \ll 1\right|_{}(i=0,1,2)$.
With Eq.~(\ref{eq_SH}) 
one finds up to order $\mathcal{O}(\epsilon_i)$
\begin{eqnarray}\label{myKG1}
 \boxempty \sqrt{\rho} = 
 \frac{\sqrt{\rho}}{\hbar^2}
 \left[\frac{\hbar^2 G_D}{3}
 (\partial^\mu \tilde{S})(\partial_\mu \tilde{S})\quad\quad\quad\quad
 \right.\quad \quad\\ \nonumber\left.
\quad
- \frac{\hbar^2}{6}
\left( 5  \Lambda
- 3  \hat{R}^{\alpha}_\alpha
 \right)\right] 
 \quad.
\end{eqnarray}
The second to last term of Eq.~(\ref{eq_dgl2c}) 
vanished, since it is proportional to the trace
of $\gamma_{\mu \nu}$,
and the last term vanished because it is of higher order in $\epsilon_i$.
A comparison of Eq.~(\ref{myKG1})
with the complex Klein-Gordon
equation~(\ref{eq_KG1}) 
shows that both equations
are the same if two identifications are made.
The first identification relates the Hamilton principal function $\tilde{S}$ to
the quantum phase $S_Q$ by
\be\label{identifyS}
 S_{Q}\equiv 
\hbar \sqrt{\frac{G_D}{3}} 
 \tilde{S} \quad.
\ee
The second identification relates
the mass $m$ to the functions 
$\hat{R}^\mu_\mu$ and $\Lambda$ by
\be\label{identifym0}
m^2\equiv  
\frac{\hbar^2}{6 }
\left( 5  \Lambda
- 3   \hat{R}^{\alpha}_\alpha
 \right)
\quad.
\ee
One can already see that such an identification makes only sense if the
scalar curvature in the four dimensional subspace 
$\hat{R}^{\mu}_\mu$ does not depend on $x_\mu$
(This is for instance the case in de-Sitter spacetimes.
One also has to check that a solution of Eq.~(\ref{myKG1}) 
allows to solve all the twenty-five equations~(\ref{loong1}-\ref{loong3})
in the $\epsilon_i$ expansion.
The first check is to compare Eq.~(\ref{myKG1})
with Eq.~(\ref{loong1}). It turns out that 
both equations are only consistent if
\be\label{cond00Tr}
\hat{R}^\mu_\mu=\Lambda\quad.
\ee
Consequently, the mass identification~(\ref{identifym0})
simplifies to
\be\label{identifym}
m^2\equiv  
\frac{\hbar^2}{3 }
  \Lambda
\quad.
\ee
This shows that a mass term can only be defined
if the four dimensional background is curved
due to a non zero cosmological constant $\Lambda$.
A similar result was previously obtained for
massive gravitons in four dimensional space-time \cite{Novello:2003ek}.
In the case of an exatly flat Minkowski background
the mass term of this toy model is zero.
Apart from the vanishing mass, an exactly flat Minkowski background
is a solution of the toy model.
The remaining twenty-four checks will be postponed to 
the  subtopic ``Is this theory consistent with all Einstein- and
conservation equations?'' in the discussion.
One also sees that for matching several different masses ($m_j$)
one has to impose several constants ($\Lambda_j$).

Since Eq.~(\ref{eq_KG1}) was found,
the next step is to derive the 
second Klein-Gordon equation~(\ref{eq_KG2}).
For this we consider 
the covariant conservation law for the 
energy-momentum tensor $T_{AB}$
\begin{eqnarray}\label{eq_cont0}
0=&\nabla_B \left(  T_{A}^B\right)=
\partial_B  T_{A}^B  - \Gamma _{AB}^D  T_{D}^B
+ \Gamma _{BD}^B  T_{A}^D\quad. 
\end{eqnarray}
Now we take the $0$ component of this equation and
apply the definitions~(\ref{eq_SH}, \ref{eq_gExpand})
\begin{eqnarray}\label{eq_cont1}
0&=& \left[
\frac{\epsilon_1}{\sqrt{\rho}} \partial_\mu\left( 
\rho \partial^\mu(\tilde{S}+\epsilon_1 \rho \beta)\right)\beta \right.\\ \nonumber
&&-\frac{\epsilon_2^2}{2}\hat{g}^{\mu \alpha}\dot{\gamma}_{\nu \alpha}
(\partial_\mu S_H)(\partial^\nu S_H)\\ \nonumber
&&+\frac{\epsilon_2^2 \epsilon_1^2}{2} \hat{g}^{\alpha \beta}\dot{\gamma}_{\alpha \beta}
\frac{\beta^2}{\alpha_0^2}
+\epsilon_0^2 \partial_0\left(\frac{\beta^2}{\alpha_0^2}\right)\\ \nonumber
&&\left. +\frac{\epsilon_1}{2}\hat{g}^{\alpha \beta} \gamma_{\alpha \beta,\, \nu}
\sqrt{\rho}(\partial^\nu S_H) \beta+{\mathcal{O}}(\epsilon_i^3)\right]
\quad. 
\end{eqnarray}
Only the first and the last term survive in a first order 
$\epsilon_i$  expansion and the
continuity equation simplifies to
\begin{eqnarray}\label{myKG2}
0&=& \partial_\mu\left( \sqrt{-\hat{g}}
\rho (\hat{g}^{\mu \nu}\partial_\nu \tilde{S})\right)
+\mathcal{O}(\epsilon_{i})
\quad.
\end{eqnarray}
With the identification~(\ref{identifyS})
this is exactly the second Klein-Gordon equation
(\ref{eq_KG2}).
Up to now we have used the classical but $\bar{t}$ dependent higher
dimensional Einstein equations~(\ref{eq_Einstein}) and
the covariant continuity equation~(\ref{eq_cont0}) and expanded
them for small perturbations around the $\bar{t}$-independent
solution.
From this we have 
found two matching rules~(\ref{identifyS}, \ref{identifym}) such that
the complex Klein-Gordon equations in a curved
spacetime~(\ref{eq_KG1}, \ref{eq_KG2})
can be exactly identified 
with the classical equations~(\ref{myKG1}, \ref{myKG2}).

\section{Critical points}

We will now try
to point out some checks, criticism, and limitations of the idea and
the derivation presented here.

{\bf{Is this quantum mechanics?}}\\
No, finding a dual to the complex Klein-Gordon
equation does not mean that one has automatically
derived quantum mechanics. To make such a statement one would
have to construct a self-consistent philosophical and
mathematical theory like the pilot wave theory of Bohm
\cite{Bohm:1951,Bohm:1951b}.
Such a construction goes beyond the scope of this paper.

{\bf{Why is the second time not visible?}}\\
The easiest
way to explain this, is with a compactification
of the hidden time such that
$\bar{t}=\bar{t}+\bar{T}$. As long the ``radius'' is just short enough,
this would only lead to violations of Lorentz invariance on
the small scale $\bar{T}$. A similar construction
for an extra but compact time variable (imaginary or not)
can be found in \cite{Damgaard:1987rr,Burinskii:2007hb}.
Thus there are several viable theoretical
models that allow and have extra time dimensions.
Within those models there
exist however experimental bounds 
on the size of the additional time
dimension $\bar{T}$ \cite{Dvali:1999hn,Bars:2006dy}.

{\bf{What is the interpretation of the higher dimensional metric?}}\\
The strongest assumption~(\ref{Ansatz0}) for the higher dimensional metric $g_{AB}$
is that $g_{00}\sim \rho$.
This can be justified by a classical
probability argument: 
The idea of this new approach becomes most clear
for  a system that does not change in the observable time
direction $t$.
In a deterministic system, the position $\vec{x}$ and velocity $\vec{v}=d\vec{x}/d\tau$
of the particle are known as soon as the initial position $\vec{x}_0$, the
initial velocity $\vec{v}_0$, and the propagation time $\tau$ are known.
Not knowing the initial conditions one has to deal with a probability
density $f(\vec{x}_0,\vec{v}_0,\tau)$. Without loss of generality we choose
some fixed starting point
which leaves a probability density $g(\vec{v}_0,\tau)$. In a deterministic
system the initial velocity can be calculated
as a function of the actual velocity and the time $\vec{v}_0=\vec{v}_0(v_{\vec{x}}(\tau),\tau)$.
Therefore, the whole system can also be described
by a statistical probability density $h(\vec{v}_x(\tau))$.
The probability $\rho(\vec{x})$
of finding the particle at a point $\vec{x}$ will than be obtained by
an integration over the proper time variable
\be
\rho(\vec{x})=\int d\tau h(\vec{v}_{\vec{x}}(\tau),\tau)\quad.
\ee
The key observation, which relates
a probability to the extra time variable, is:
The faster a particle moves at a certain point $\vec{x}$ the smaller
is the probability of finding the particle at this point
$h(\vec{v}_{\vec{x}}(\tau),\tau)\sim  1/\vec{v}_{\vec{x}}(\tau)$ and hence
\be\label{rhoV}
\rho(x)\sim \int d\tau \frac{1}{v_x(\tau)}
=\int d\tau \frac{1}{\frac{dx}{d\tau}}\quad.
\ee
This probability is directly related to the $00$ component of the
metric $g_{AB}$.
We already assumed that the particle moves mostly in parallel 
to the additional time coordinate ($d\bar{t}/d\tau \gg dx/d\tau > dt/d\tau$).
In this limit the differential of the proper time $\tau$ can be approximated
by the differential of the additional time variable $d\tau\approx d\bar{t}\;  \sqrt{ g_{00}}$.
Under the assumption that some part of 
From the integral in Eq.~(\ref{rhoV}) one finds
\be
\rho(x)\sim\int d\bar{t}  \frac{1}{\frac{dx}{d\bar{t}}} \;\,  g_{00}\quad.
\ee
This shows that a statistical interpretation of an unobservable additional
time leads naturally to the metric ansatz~(\ref{Ansatz0}).

{\bf{
Is this theory consistent with all Einstein- and
conservation equations?}}\\
Since we only worked with a subset
of the Einstein equations we have to check whether
this can lead to inconsistencies with the full set.
From  comparing Eq.~(\ref{myKG1})
with Eq.~(\ref{loong1}) we already found
that $\hat{R}^\mu_\mu=\hat{\Lambda}$.
The same condition is found when comparing
Eq.~(\ref{myKG1}) with the trace of Eq.~(\ref{loong3}).
Because of this condition one can
always decompose the four dimensional Ricci tensor as
\be\label{Rhat}
\hat{R}_{\mu \nu}=n_1 \hat{g}_{\mu \nu}
+G_D p_\mu p_\nu \quad,
\ee
where the four vector $p_\mu$ and the number
$n_1$ do not depend on $x_\mu$.
Please note that we do not know the solution to Eq.~(\ref{Rhat})
but we assume that it exists.
In the $\epsilon_i$ expansion the eight equations~($\ref{loong2}$) 
do not create any new conditions,
since the LHS is of order $\epsilon_i^2$ and the
RHS is composed of a vanishing $\bar{t}$ boundary term 
plus a term which is also of higher order in $\epsilon_i$.
In the same scheme and
after using Eqs.~(\ref{cond00Tr}, \ref{Rhat}) the remaining nine independent equations 
in~(\ref{loong3}) read
\begin{eqnarray}\label{loong3b}
\int d\bar{t} \frac{(\sqrt{\rho})_{;\delta;\beta}}{\sqrt{\rho}}
=-\int d\bar{t}\,\left[G_D \left( 
(\partial_\delta \tilde{S}) (\partial_\beta\tilde{S}) -p_\delta p_\beta
\right)\right.\\ \nonumber
-\frac{g_{\delta \beta}}{3} (G_D (\partial^\mu \tilde{S}) (\partial_\mu \tilde{S})- \Lambda)
+\frac{g_{\delta \beta}}{4}\left.(G_D p^\mu p_\mu-\Lambda)
\right]\quad.
\end{eqnarray}
It is hard to prove in general that those
equations and the Klein-Gordon equations
can always be consistently solved. However, we can show that
this is the case for the most simple solution of the Klein-Gordon
equation. The most simple solution of 
the first Klein-Gordon equation~(\ref{myKG1}) is: $\rho=$const,
$\tilde{S}_{,\mu}\equiv p_\mu$, and $p^\mu p_\mu=\Lambda/G_D$.
For this solution the nine independent  equations~(\ref{loong3b})
are obviously fulfilled since both sides of the equation
vanish. A comparison with Eq.~(\ref{cond00Tr}) further shows
that $n_1=0$.
Thus, we have shown that, at least in 
the discussed special case, all twenty-five Einstein equations
can be fulfilled consistently in our theory. 
As last consistency check we have to study the
four remaining continuity equations from
Eq.~(\ref{eq_cont0})
\begin{eqnarray}
\nabla_B \left(T_\mu^B \right)=\partial_\delta
\left(g^{\delta \nu}\tilde{S}_{,\mu} \tilde{S}_{,\nu} \right)
-\hat{\Gamma}^\delta_{\mu \nu}\left(g^{\gamma \nu}\tilde{S}_{,\gamma} \tilde{S}_{,\delta} \right)\quad\quad \\ \nonumber
+\hat{\Gamma}^\nu_{\delta \nu}\left(g^{\gamma \delta}\tilde{S}_{,\mu} \tilde{S}_{,\gamma} \right)+\frac{\rho_{,\nu}}{2 \rho}
\left(g^{\gamma \nu}\tilde{S}_{,\gamma} \tilde{S}_{,\mu}  \right)
+\mathcal{O}(\epsilon_i)\quad.
\end{eqnarray}
For the above solution of the Klein-Gordon equation
this is the conservation law for the four momentum $p_\mu \sim \tilde{S}_{,\mu}$.
Therefore, we have checked the consistency
of an exemplary solution of our theory with all its thirty
equations~(\ref{loong1}-\ref{loong3},\ref{eq_cont0}).
Because of the condition~(\ref{Rhat})
the Klein-Gordon equation on a flat Minkowski
background can only result from our theory
in an asymptotic region of spacetime.
Since energy always curves spacetime,
this what one would expect also from
the standard interpretation of quantum mechanics
and general relativity.

{\bf{What is the classical limit?}}\\
The construction with its higher dimensional
energy-momentum tensor $T_{AB}$ and the higher dimensional metric
$g_{AB}$ should (in the limit of large
distancescales and large timescales) be consistent with the
usual four dimensional classical general relativity.
In this limit the $\rho$ dependence and the terms
proportional to $\epsilon_i$ can be omitted.
However, a different normalization in 
the metric expansion~(\ref{eq_gExpand}) 
is useful such that $\hat{g}_{\mu \nu}\rightarrow 3 g^0_{\mu \nu}/2$.
Now Eq.~(\ref{loong3})
simplifies to
\be
\hat{R}_{\mu \nu}=G_D\left(
\tilde{S}_{,\mu}\tilde{S}_{,\nu}
-\frac{g^0_{\mu \nu}}{2} \tilde{S}^{,\alpha} \tilde{S}_{,\alpha}\right)
+\frac{g^0_{\mu \nu} \Lambda}{2 }\quad.
\ee
For an infinitely small normalizing volume at the
position of the particle trajectory 
we denote the density of the energy momentum tensor as
$\tilde{S}_{,\mu}\tilde{S}_{,\nu} \rightarrow \tilde{p}_\mu \tilde{p}_\mu 
\delta^3(\vec{x}-\vec{x}_n)/E$.
Thus, we found the four dimensional Einstein field equations for a free
particle \cite{Weinberg:1972}
\be\label{classicalLimit}
\hat{R}_{\mu \nu}=\frac{G_D}{E}
\left(\tilde{p}_{\mu}\tilde{p}_{\nu}+\frac{ g^0_{\mu \nu}}{2}\tilde{p}^2
\right)
\delta^3(\vec{x}-\vec{x}_n)
+\frac{g^0_{\mu \nu}}{2}\Lambda\quad.
\ee
Please note that in contrast to \cite{Arkani-Hamed:1998rs,Antoniadis:1998ig} the
gravitating matter is free to propagate into all space and time dimensions
and therefore the higher dimensional gravitational
coupling $G_D$ is equal to the Newton constant $G_N$.
This shows that classical general relativity can be obtained
from the ansatz~(\ref{eq_Einstein}) after integrating out $\bar{t}$ and taking the
limit of large time and distance scales.

{\bf{Isn't the cosmological constant very small?}}\\
The higher dimensional cosmological term $\Lambda$
is on the one hand related to the mass in the quantum equation
and on the other hand to the four dimensional
cosmological constant $\Lambda_4$.
The observed cosmological constant $\Lambda_4$ is
a very small number \cite{PDBook} and therefore inconsistent 
with a typically much bigger particle mass in Eq.~(\ref{identifym}).
A possible way out of the mass-cosmological constant problem
might be found by considering higher orders in the $\epsilon_i$
expansion or by taking a possible $\bar{t}$ dependence of $\Lambda$
into account.
Even if the issue remains in our toy model,
this might be a good sign because also most of the conventional 
quantum theories have a problem in getting the
cosmological constant right \cite{Weinberg:1988cp}.

{\bf{Which parameters had to be engineered?}}\\
The presented mechanism only worked after demanding
Eqs.~(\ref{identifyS}, \ref{identifym}) to be true.
The first equation~(\ref{identifyS}) relates the
classical principal functional $\tilde{S}$
to the phase of the quantum wave function $S_Q$.
The second equation~(\ref{identifym})
relates the constant $\Lambda$
to the mass of the quantum particle. 

{\bf{What about spinors?}}\\
One of the great successes of quantum mechanics
was the prediction and explanation of a particle with 
a Land{\'e} factor ($g=2$)
due to the Dirac equation.
It will be a challenging  task to find the analog of the
Dirac equation in our picture.
However, it has been shown by \cite{Carter:1968ks} that 
the $g=2$ factor is also present in a Kerr-Newman spacetime. 
Later, \cite{Burinskii:2007hb} found out how to relate 
the electro magnetic field of a Dirac electron
to the surrounding field of the  Kerr-Newman
solution.
Those progresses indicate that there could be a way to explain
spin $1/2$ with the help of classical general relativity.

{\bf{Quantum field theory?}}\\
Questions concerning the formulation of the second quantization
in quantum field theory, interacting field theory,
or Yang Mills theories in this picture are hopefully subject
to future studies.
%
%
%
%
%

\section{Summary}

In this paper we study the higher dimensional
Einstein equations~(\ref{eq_Einstein}) 
and the higher dimensional continuity equation~(\ref{eq_cont0})
with one additional time dimension $\bar{t}$.
For the higher dimensional 
metric we make the ansatz~(\ref{Ansatz0}),
and for the higher dimensional principal function we
make the ansatz~(\ref{eq_SH}).
Then we make an expansion in the parameter $\epsilon_i$
to ensure a small $\bar{t}$ dependence.
After defining the
quantum phase (\ref{identifyS})
and the quantum mass (\ref{identifym})
we find the structure of the Klein-Gordon equation 
in curved spacetime
\begin{eqnarray}\nonumber
(\ref{myKG1}):
 \quad \; \boxempty \sqrt{\rho} &=& 
 \frac{\sqrt{\rho}}{\hbar^2}
 \left[
 (\partial^\mu S_Q)(\partial_\mu S_Q)-m^2\right] +\mathcal{O}(\epsilon_{i})\quad,
 \\ \nonumber
(\ref{myKG2}):\quad \quad \quad
0&=& \partial_\mu\left( 
\sqrt{-\hat{g}} \hat{g}^{\mu \nu} \rho (\partial_\nu S_Q)\right)
+\mathcal{O}(\epsilon_{i})
\quad.
\end{eqnarray}

Thus, by constructing this toy model
we explicitly showed that it is possible
to find a gravitational dual to the complex
Klein-Gordon equation in curved space-time.
This achievement of the model is accompanied by a number
of possible issues which have to be further understood.
First, the weak $\bar{t}$ dependence of the functions $S_H$ and
$\bar{\alpha}$ is just an ansatz and not a necessity.
Second, except of the case with $m=0$ we could not
find a solution of
all twentyfive Einstein equations. 
Third, having a gravitational dual to the
complex Klein-Gordon equation
does not mean that one
also has a dual for relativistic quantum mechanics.
However, the toy model is one of
several (very different) theories 
\cite{Damgaard:1987rr,Maldacena:1997re,Witten:1998qj,Hull:1999mr,Weinstein:2007} 
that make a formal connection
between the quantum structure of nature and an introduction
of an additional time dimension.
Our conclusion is that this connection could be more
than a pure mathematical peculiarity, it might actually reflect
the structure of spacetime.\\ \\
Many thanks to 
Jorge Noronha, Silke Weinfurtner, Goirgio Torrieri, 
Sabine Hossenfelder, Martin Kober, Nan Su, and Basil Sad
for their comments and remarks.


\begin{thebibliography}{10}

\bibitem{Bohm:1951}
D.~Bohm,
\newblock Phys. Rev. {\bf 85}, 166 (1951).

\bibitem{Bohm:1951b}
D.~Bohm,
\newblock Phys. Rev. {\bf 85}, 180 (1951).

\bibitem{Klein:1926a}
O.~Klein,
\newblock Z. f. Physik {\bf 37}, 895 (1926).

\bibitem{Hooft:1999gk}
G.~'t~Hooft,
\newblock Class. Quant. Grav. {\bf 16}, 3263 (1999), gr-qc/9903084.

\bibitem{Biro:2001dh}
T.~S. Biro, S.~G. Matinyan, and B.~Muller,
\newblock Found. Phys. Lett. {\bf 14}, 471 (2001), hep-th/0105279.

\bibitem{Nikolic:2006az}
H.~Nikolic,
\newblock Found. Phys. {\bf 37}, 1563 (2007), quant-ph/0609163.

\bibitem{Dvali:1999hn}
G.~R. Dvali, G.~Gabadadze, and G.~Senjanovic,
\newblock (1999), hep-ph/9910207.

\bibitem{Bars:2000qm}
  I.~Bars,
  Class.\ Quant.\ Grav.\  {\bf 18} (2001) 3113
  [arXiv:hep-th/0008164].

\bibitem{Bars:2001ma}
  I.~Bars and S.~J.~Rey,
  Phys.\ Rev.\  D {\bf 64} (2001) 046005
  [arXiv:hep-th/0104135].

\bibitem{Villanueva:2005up}
  V.~M.~Villanueva, J.~A.~Nieto, L.~Ruiz and J.~Silvas,
  J.\ Phys.\ A  {\bf 38} (2005) 7183
  [arXiv:hep-th/0503093].


\bibitem{Bars:2006dy}
I.~Bars,
\newblock Phys. Rev. {\bf D74}, 085019 (2006), hep-th/0606045.

\bibitem{Seahra:2001bx}
S.~S. Seahra and P.~S. Wesson,
\newblock Gen. Rel. Grav. {\bf 33}, 1731 (2001), gr-qc/0105041.

\bibitem{Hooft:1993gx}
G.~'t~Hooft,
\newblock (1993), gr-qc/9310026.

\bibitem{Susskind:1994vu}
L.~Susskind,
\newblock J. Math. Phys. {\bf 36}, 6377 (1995), hep-th/9409089.

\bibitem{Maldacena:1997re}
J.~M. Maldacena,
\newblock Adv. Theor. Math. Phys. {\bf 2}, 231 (1998), hep-th/9711200.

\bibitem{Witten:1998qj}
E.~Witten,
\newblock Adv. Theor. Math. Phys. {\bf 2}, 253 (1998), hep-th/9802150.

\bibitem{Hull:1999mr}
C.~M. Hull,
\newblock (1999), hep-th/9911080.

\bibitem{Weinstein:2007}
S.~Weinstein,
\newblock Physicsworld {\bf Sept.}, 18 (2007).

\bibitem{Damgaard:1987rr}
P.~H. Damgaard and H.~Huffel,
\newblock Phys. Rept. {\bf 152}, 227 (1987).

\bibitem{Birrell}
N.~D. Birrell and P.~C.~W. Davies,
\newblock {Cambridge {U}niversity {P}ress; ISBN: 0 521 27858 9} , 43+ (1982).

\bibitem{Kaluza:1921}
T.~Kaluza,
\newblock Sitzungsber. Preuss. Akad. Wiss. Phys. math. {\bf Klasse}, 996
  (1921).

\bibitem{Klein:1926b}
O.~Klein,
\newblock Nature {\bf 118}, 516 (1926).

\bibitem{Wesson:2000}
S.~Wesson, Paul,
\newblock {\em Space - Time - Matter} (World Scientific Publishing Co. Pte.
  Ldt.; ISBN 981-02-3588-7, New York 07661, 2000).

\bibitem{Hamilton:1833}
W.~Hamilton,
\newblock Dublin University Review pp. , 795 (1833).

\bibitem{Landau:1975}
L.~Landau and L.~Lifshitz,
\newblock Elsevier, Amsterdam; ISBN 0750628960  (1975).


\bibitem{Bernard:1977pq}
C.~W. Bernard and A.~Duncan,
\newblock Ann. Phys. {\bf 107}, 201 (1977).


\bibitem{Novello:2003ek}
  M.~Novello and R.~P.~Neves,
  Class.\ Quant.\ Grav.\  {\bf 20} (2003) L67.


\bibitem{Burinskii:2007hb}
A.~Burinskii,
\newblock (2007), arXiv:0712.0577 [hep-th].

\bibitem{Bell:1964kc}
J.~S. Bell,
\newblock Physics {\bf 1}, 195 (1964).

\bibitem{Baym:1990QM}
G.~Baym,
\newblock {\em Lectures on quantum mechanics} (Perseus Books Publishing, New
  York, 1990).

\bibitem{Wesson:2002zm}
P.~S. Wesson,
\newblock (2002), gr-qc/0205117.

\bibitem{Weinberg:1972}
S.~Weinberg,
\newblock {\em Gravitation and Cosmology} (ISBN 0-471-92567-5, 1972).

\bibitem{Arkani-Hamed:1998rs}
N.~Arkani-Hamed, S.~Dimopoulos, and G.~R. Dvali,
\newblock Phys. Lett. {\bf B429}, 263 (1998), hep-ph/9803315.

\bibitem{Antoniadis:1998ig}
I.~Antoniadis, N.~Arkani-Hamed, S.~Dimopoulos, and G.~R. Dvali,
\newblock Phys. Lett. {\bf B436}, 257 (1998), hep-ph/9804398.

\bibitem{PDBook}
W.-M. e.~a. {Yao},
\newblock {Journal of Physics G} {\bf 33}, 1+ (2006).

\bibitem{Weinberg:1988cp}
S.~Weinberg,
\newblock Rev. Mod. Phys. {\bf 61}, 1 (1989).

\bibitem{Carter:1968ks}
B.~Carter,
\newblock Commun. Math. Phys. {\bf 10}, 280 (1968).

\end{thebibliography}

\end{document}